%% file: main.tex
\documentclass[conference,a4paper]{IEEEtran}
\IEEEoverridecommandlockouts

\usepackage{amsmath}
\usepackage{bm}
\usepackage{mdframed}
\usepackage{amsthm}

\usepackage[utf8]{inputenc}
\usepackage{graphicx}
\usepackage[colorinlistoftodos]{todonotes}
\usepackage{amssymb}
\usepackage{longtable}
\usepackage{nohyperref}
\hypersetup{bookmarks=false}
\usepackage[noadjust]{cite}
\allowdisplaybreaks
\usepackage{lineno}
\usepackage{lipsum}
\usepackage{multicol}
\usepackage{subcaption}

\usepackage{multirow}
\usepackage{array}

\usepackage{xcolor}
\usepackage[vlined,linesnumbered,ruled,resetcount]{algorithm2e}
\usepackage{bbm}

\SetKwInput{KwInput}{Input}                
\SetKwInput{KwOutput}{Output}              
\SetKwInput{KwInitialize}{Initialize}                
\SetKwInput{myproc}{Procedure}{}{}

\usepackage{booktabs}
\usepackage{caption}
\usepackage{float}
\usepackage{capt-of}

\usepackage{arydshln}
\renewcommand{\baselinestretch}{0.98}

\title{Deep Reinforcement Based Optimization of Function Splitting in Virtualized Radio Access Networks}

\author{\IEEEauthorblockN{Fahri Wisnu Murti, Samad Ali, and Matti Latva-aho}\\
\vspace{-5mm}
\IEEEauthorblockA{
	Centre for Wireless Communications, University of Oulu, Finland\\
	Email: fahri.murti@oulu.fi, samad.ali@oulu.fi, matti.latva-aho@oulu.fi
	}
\thanks{{This work was supported by the Academy of Finland 6Genesis Flagship (grant no. 318927).}}%
}

\begin{document}

\maketitle

\begin{abstract}
Virtualized Radio Access Network (vRAN) is one of the key enablers of future wireless networks as it brings the agility to the radio access network (RAN) architecture and offers degrees of design freedom. Yet, it also creates a challenging problem on how to design the functional split configuration. In this paper, a deep reinforcement learning approach is proposed to optimize function splitting in vRAN. A learning paradigm is developed that optimizes the location of functions in the RAN. These functions can be placed either at a central/cloud unit (CU) or a distributed unit (DU). This problem is formulated as constrained neural combinatorial reinforcement learning to minimize the total network cost. In this solution, a policy gradient method with Lagrangian relaxation is applied that uses a stacked long short-term memory (LSTM) neural network architecture to approximate the policy. Then, a sampling technique with a temperature hyperparameter is applied for the inference process. The results show that our proposed solution can learn the optimal function split decision and solve the problem with a $0.4\%$ optimality gap. Moreover, our method can reduce the cost by up to $320\%$ compared to a distributed-RAN (D-RAN). We also conclude that altering the traffic load and routing cost does not significantly degrade the optimality performance.
\end{abstract}


\IEEEpeerreviewmaketitle


\vspace{-2mm}
\input{Introduction.tex}


\vspace{-2mm}
\input{System_model.tex}

\vspace{-2mm}
\input{Problem_formulation.tex}

\vspace{-2mm}
\input{Algorithms.tex}


\vspace{-2mm}
\input{Simulations.tex}

\vspace{-2mm}
\input{Conclusions.tex}


\bibliographystyle{IEEEtran}
\bibliography{IEEEabrv,ref-vran-01}

\end{document}

%% file: Introduction.tex
\section{Introduction}
\vspace{-1mm}

The increase in mobile data traffic of emerging applications with stringent requirements has driven the efforts to re-design the radio access networks (RANs). To this end, there have been systematic works in standardization bodies to adopt \textit{softwarization} and \textit{virtualization} of RAN architecture \cite{split_3gpp,split_3gpp_rel16,smallcell}. Centralized/Cloud RAN (C-RAN) has become a promising
solution to enable the deployment of low-cost and high-performance systems by pooling the base station (BS) functions in a central server which is also known as Cloud Unit (CU). However, C-RAN is difficult to implement for many reasons. For example, it requires a low-latency and high capacity fronthaul which are often not available in current RANs and costly to build from scratch. Such challenges motivate the shift of rigid C-RAN to flexible architectures where a subset of BS functions is hosted at CU, and the other functions are at distributed units (DUs).

The term \textit{virtualized RAN} (vRAN) is used to describe an architecture that allows to deploy different \textit{functional split} for each BS \cite{nokia_anyhaul}. However, selecting a functional split configuration (which functions to deploy at CU and which functions at the DU) for each BS creates an intricate problem. Each split differs in delay requirements, initiates different computation loads for CU and DU, and creates different data flows. Additionally, there is a consent that the original design using an evolved Common Public Radio Interface (eCPRI)/CPRI should be replaced by integrated fronthaul/backhaul (Crosshaul) based packet-switch (shared) network which is more cost-efficient \cite{wizhaul_andres}. As a result, functional split requires a careful design, especially in networks with limited capacity. These issues testify that optimizing the split configuration is vital, although it may increase RAN management's complexity.  


3GPP \cite{split_3gpp,split_3gpp_rel16} and a seminal white paper \cite{smallcell} have defined the detail vRAN split specifications. Although the authors in  \cite{function_split_survey} have discussed the gains and requirements of vRAN split, there are still limited works on the optimization issues. Energy consumption for various splits has been evaluated in \cite{apt-ran}, then the authors have proposed an optimization model over different splits. The authors in \cite{wizhaul_andres} studied optimizing the centralization degree of C-RAN/vRAN over Crosshaul. Follow up works, \cite{fluidran_andres} and \cite{vranmec_andres} offered optimal solution of minimizing total cost for integration vRAN with Mobile Edge Computing (MEC). Then, \cite{vran_murti1} proposed an optimized multi-cloud vRAN framework with balancing its centralization \cite{vran_murti2}. However, the mentioned works above rely on mathematical optimization techniques that often have a slow convergence rate and exponential complexity for finding the optimal solution, particularly in large networks. Moreover, these optimization-based approaches heavily rely on expert knowledge for formulating each problem mathematically, which may be insufficient in practice \cite{mlcombi_yoshua}. Besides, the above problems are often combinatorial and difficult to solve. 


In operational research, \textit{machine learning} (ML) approaches have spurred to address combinatorial optimization problems without much handcrafted engineering and heuristic algorithm design \cite{mlcombi_yoshua}. 
For instance, the authors in \cite{pointer} proposed a supervised learning based on Pointer Networks (Ptr-Nets); but, it requires access to optimal labels that may not be possible in many problems. Bello \emph{et al} proposed Neural Combinatorial Optimization (NCO) with an end-to-end approach via neural network and reinforcement learning (RL) to tackle this limitation \cite{neural_bello}. Further studies have shown that this approach successfully solves combinatorial problems, e.g., 0-1 Knapsack, Travelling Salesman \cite{neural_bello}, device placement \cite{device_azalia}, with a near-optimal solution and fast execution time. 

Inspired by \cite{neural_bello}, Jiang \emph{et al.} proposed RL with Multi-Pointer networks (Mptr-Net) to solve the offloading problem in MEC and showed that their approach attained more than $98\%$ of optimality \cite{neural_jiang}. The authors in \cite{vnf_drl_solozabal} also proposed a deep RL approach with a sequence-to-sequence model to solve the virtual network function (VNF) placement problem and aimed to minimize the power consumption. Recent work \cite{vranai_journal} proposed a vrAIn framework, a deep RL approach for dynamic computing and radio resources control in the vRAN system. Although such approaches are promising in solving complex combinatorial problems for zero-touch optimization in wireless network \cite{ali20206g,ml_commnetwork}, there is still no prior work to employ it for functional split optimization in vRAN.


Our goal is to develop a deep RL framework for zero-touch optimization of split configuration for each BS in a vRAN system. First, we formulate and present the vRAN splitting as an optimization problem to provide a better understanding of its objective and constraints. The resulting problem is an NP-hard with prohibitive complexity for a large network and real-time execution. Motivated by \cite{neural_bello,vnf_drl_solozabal}, we formulate the problem above as constrained neural combinatorial reinforcement learning and develop a solution approach, namely \texttt{DRLT-vRAN}. It is worth noting that our approach requires minimal handcrafted engineering. It does not need to know the vRAN split problem mathematically, e.g., mathematical optimization-based approaches \cite{fluidran_andres, vranmec_andres, vran_murti1,vran_murti2}, or direct access to the optimal labeled data, e.g., supervised learning \cite{vran_drl}. Instead, it learns from interaction with the environment that expects to receive the reward (total network cost) signal and Penalization (constraints violation). To the best of our knowledge, this is the first work using constrained deep RL paradigm to solve the functional split optimization in the vRAN system.

\texttt{DRLT-vRAN} aims to approximate the policy that optimizes the split configuration. It is tailored from a Policy Gradient \cite{rl_sutton} with Lagrangian relaxation method \cite{reward_constraint,vnf_drl_solozabal} that uses a neural network architecture. The neural network architecture is a sequence-to-sequence model with attention mechanism, formed by encoder-decoder design, and based on stacked Long Short Term Memory (LSTM) networks \cite{seq2seq,attention_bahdanau}. Also, a baseline estimator is separately trained in an auxiliary network to improve the policy further. Then, we use a searching strategy, which is a sampling technique with temperature hyperparameter, for the inference process \cite{neural_bello}. 



We evaluate our approach in a synthetic network generated by the Waxman algorithm that highly represents a backhaul network \cite{waxman}. The used system parameters are from a measurement-based 3GPP-compliant system model\cite{vran_murti1,vran_murti2}. To assess our approach's effectiveness, we compare it to the optimal value obtained from a Phyton-MIP solver\footnote{A mixed-integer programming solver (https://www.python-mip.com/)}. Following our evalutions, \texttt{DRLT-vRAN} successfully learns the optimal function split decision, solves the problem with less than a $0.4\%$ optimality gap, and saves the total network cost to $320\%$ of D-RAN. Additionally, altering the traffic load and routing cost does not significantly degrade the optimality performance.

The rest of this paper is organized as follows. Section \ref{sec:model} presents the background and system model of vRAN. We formalize the vRAN split problem mathematically in Section \ref{sec:problem}. Section \ref{sec:solution} describes our solution approach, \texttt{DRLT-vRAN}. We discuss our experiment results in Section \ref{sec:results} and finally conclude our work in Section \ref{sec:conclusion}.

%% file: System_model.tex
\section{System Model} \label{sec:model}
\vspace{-1mm}

\begin{figure}[t!]
	\centering
	\includegraphics[width=0.42 \textwidth]{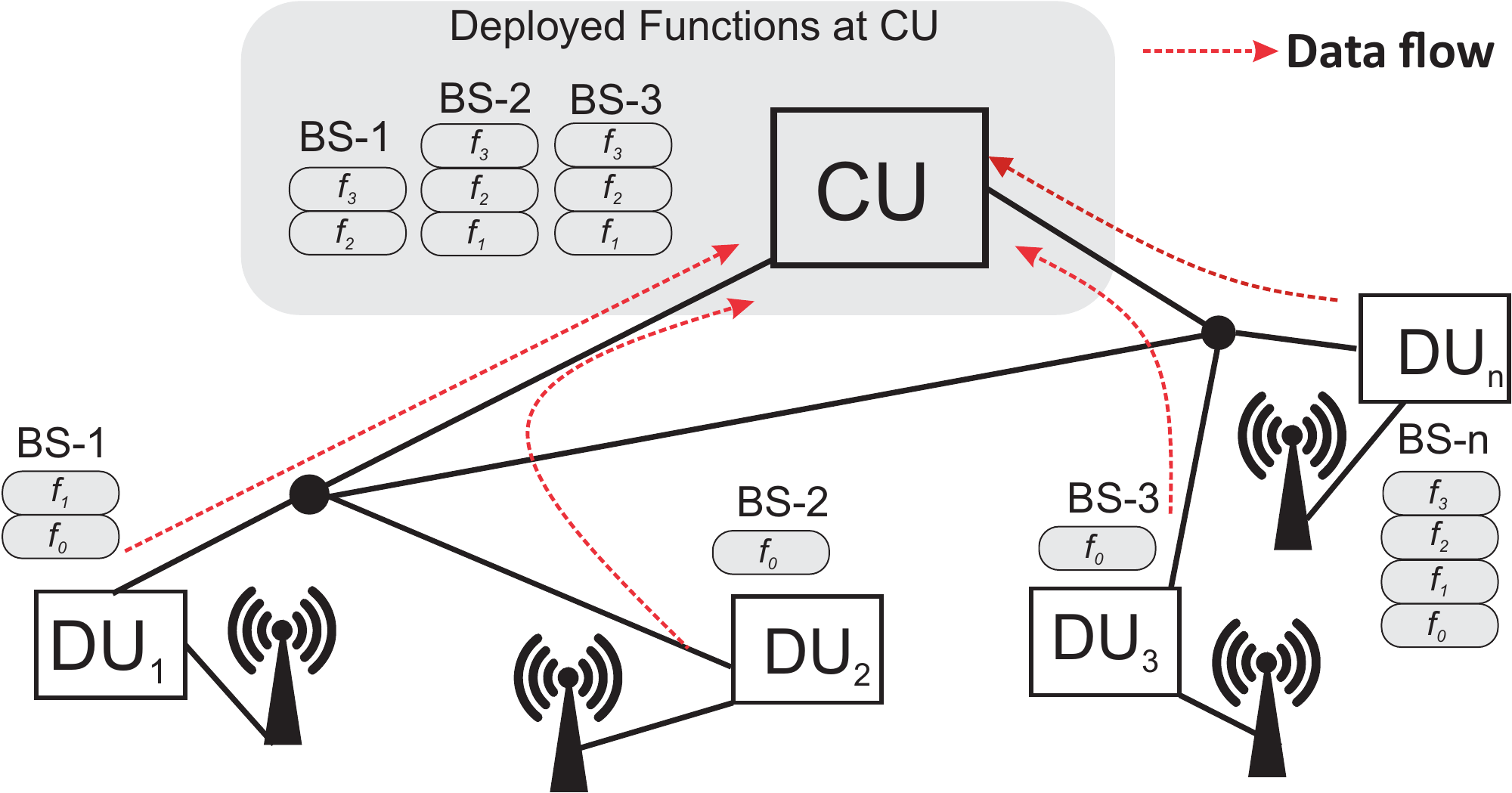}   
	\caption{\small vRAN over Crosshaul. It has many degrees of design freedom by possibly hosting BS functions at CU or DU.}
	\label{fig:vran}
	\vspace{-5mm}
\end{figure}

\textbf{Background.} In C-RAN, all BS functions are at Base Band Unit (BBU), except RF layers at Radio Unit (RU).  In the development, BBU functions are decoupled into CU and DU \cite{split_3gpp_rel16}. Hence, a BS consists of CU, DU, and RU. Fig \ref{fig:vran} shows that a CU is typically a bigger server and placed in a central location, while DU is smaller and located near (or co-located) with RU. Table \ref{table:splits} describes the particular vRAN split options and their requirements. 

Our model refers to the standardization of 3GPP \cite{split_3gpp,split_3gpp_rel16} and seminal white paper \cite{smallcell}, where each split has a different performance gain \cite{vran_murti2,function_split_survey}. \textbf{Split 0}: All functions are at DU, except the RF layer is at RU. It is a typical D-RAN setup. \textbf{Split 1} (PDCP-RLC): RRC, PDCP, and upper layers are hosted at CU, while RLC, MAC, and PHY are at DU. It enables L3 and L2 operation at the same server. \textbf{Split 2} (MAC-PHY): MAC and upper layers are at CU; PHY at DU. It allows improvement for CoMP by centralized HARQ. \textbf{Split 3} (PHY-RF): All functions are at CU, except RF layers. It is a fully centralized version of vRAN, and gains power-saving and improved joint reception CoMP with uplink PHY level combining. Going from Split 1 to 3, more functions are hosted at CU. In addition to increasing network performance, a higher centralization level can lead to more cost-saving \cite{vran_murti2}. However, centralizing more functions increases the data load to be transferred to CU, going from $\lambda$ in S0 to 2.5 Gbps in S3 for each BS, and has stricter delay requirements (Table \ref{table:splits}).

\begin{table}[t!] \centering
	\begin{small}
		\begin{tabular}{@{}lll@{}}\toprule
			\textbf{Split}& \textbf{Flow (Mbps)} & \textbf{Delay Req. (ms)}  
			\\ \midrule
			\textbf{$0$} &      $\lambda$   & $30$          
			\\ \hdashline
			{\textbf{$1$}} &  {$\lambda$} & $30$ 
			\\ \hdashline
			{\textbf{$2$}} &  {$1.02\lambda+1.5$}   & $2$
			\\ \hdashline
			{$3$} &   {$2500$}   & $0.25$ \\ 
			\bottomrule
		\end{tabular}
	\end{small}
	\caption{\small Flow and delay requirements (the traffic load is $\lambda$ Mbps).}
	\label{table:splits}
	\vspace{-7mm}
\end{table}

\textbf{RAN}. We model a vRAN architecture with a graph $G = (\mathcal{I}, \mathcal{E})$ where $\mathcal{I}$ has a subsets $\mathcal{N}$ of the $N$ DUs, routers, and a CU (index $0$). Each node is connected through a link of $(i,j)$ with a set $\mathcal{E}$, and has capacity $c_{ij}$ (Mbps) each. The DU-$n$ is connected to $\{0\}$ with a single path (e.g., shortest path) $p_{n0}$; hence, we define $r_{p_{n0}}$ as the amount of data flow (Mbps) is transferred and routed through path $p_{n0}:=  \{(n,i_1), ..., (i_k,0):(i,j) \in \mathcal{E} \}$. The BS functions are deployed in servers (DU or CU) using virtual machines (VMs). Each server has a processing capacity, i.e., $H_n$ for DU-$n$ and $H_0$ for CU. Naturally, a central server has a higher computational capacity and lower processing load (cycle/Mb/s). Hence, we define that $H_0 \geq H_n$ and $\rho_{o}^c \leq \rho_{o}^d$, where $\rho_{o}^c$ and $\rho_{o}^d$ are the computational processing load in result of deploying the split configuration $o \in \{0,1,2,3\}$ at CU and DU, respectively.

\textbf{Demand $\&$ Cost}. We focus on the uplink transmission where $\lambda_{n}$ (Mbps) is the aggregate data flow of DU-$n$ to serve the users traffic;
hence, there are $N$ different flows in the network. We denote $ \bm{\alpha} = (\alpha_n, n \in \mathcal{N})$ and $\bm{\beta} = (\beta_n, n \in \mathcal{N})$ as cost for instantiating the VM (monetary units) and the computing cost (monetary units/cycle) at DU-$n$, respectively, while $\alpha_0$ and $\beta_0$ are the respective cost for CU. We also have a routing cost $\zeta_{p_{n0}}$ (monetary units/Mbps) for each path $p_{n0}$. This cost arises from the network link are leased from third parties or the expenditures of maintaining the link. 

\textbf{Problem Statement.} We have four choices of split configurations for each BS in vRAN. What is the best-deployed split configuration for each BS that minimizes the total network cost? The decision leads to interesting problems. Each configuration generates a different DU-CU data flow and has a distinct delay requirement. Executing more functions at CU is more efficient in terms of computing cost; however, it produces a higher load of crosshaul links. We also have delay processing and limited capacity for each server and network link.

%% file: Problem_formulation.tex
\section{Formalization of vRAN Split Problem} \label{sec:problem}
\vspace{-1mm}
The BS functions can be deployed at DU (for each BS) or pooled at CU according to the split configuration, see Table \ref{table:splits}. The configuration must respect to  the \emph{chain of functions} $f_0 \!\rightarrow\! f_1 \!\rightarrow\! f_2 \!\rightarrow\! f_3$. Thus, we define binary variable $x_{on}$ as the decisions for deploying split $o$  at DU-$n$. For instance, $x_{0n} = 1$ is for deploying $f_0, f_1, f_2, f_3$ (Split 0); $x_{1n} = 1$ for $f_0, f_1, f_2$ (Split 1); $x_{2n}= 1$ for $f_0, f_1$ (Split 2); or $x_{3n}= 1$ for $f_0$ (Split 3) at DU-$n$. 
We only deploy a single configuration for each BS. Therefore, the set of eligible split configuration is:
\vspace{-1mm}
\begin{align} \label{eq:setx}
\mathcal{X} =  &\Bigl\{  \bm{x}_n  \in \{0,1\} \Big| \sum_{o=0}^3 x_{on}=1 , 
\ \ \forall n \in \mathcal{N}  \Bigr\} , 
\end{align}
where $\bm{x}_n = ( x_{on}, \forall o  )$ and $\bm{x} = (\bm{x}_n, \forall n)$. The BS functions,  $f_1, f_2$ and $f_3$, are deployed in servers with VMs at each server. We have computational processing for CU and for each DU that must respect its capacity as:
\vspace{-1mm}
\begin{align} \label{eq:computing1}
\sum_{n \in \mathcal{N}} \lambda_{n} \sum_{o =0}^3    x_{on} \rho^\text{c}_o \leq \ H_0, 
\end{align}
\vspace{-1mm}
\vspace{-1mm}
\begin{align} \label{eq:computing2} \lambda_{n} \sum_{o =0}^3   x_{on} \rho^\text{d}_o \leq H_n, \ \forall n \in \mathcal{N}. 
\end{align}
%

\textbf{Data Flow $\&$ Delay.}  Variable $r_{p_{n0}}$ (Mbps) defines the amount of data flow (Mbps) are transferred through path $p_{n0}$. Hence, the flow must respect the link capacities:
\vspace{-1mm}
\begin{align} \label{eq:route1}
\sum_{n \in \mathcal{N}} r_{p_{n0}} I^{ij}_{p_{n0}} \leq c_{ij}, \ \ \forall (i,j) \in \mathcal{E},
\end{align}
where $I^{ij}_{p_{n0}} \in \{ 0,1 \}$ indicates whether the link $(i,j)$ is used by path $p_{n0}$. Assuming a single path (e.g., shortest path), the amount of data flow depending on the split options is \cite{fluidran_andres}:
\vspace{-1mm}
\begin{align} \label{eq:route2}
r_{p_{n0}} \!=\! \lambda_{n} (x_{0n} + x_{1n}) + x_{2n} (1.02 \lambda_{n} + 1.5) + 2500 x_{3n}.
\end{align}
Lets denote $d_{p_{n0}}$ is a delay incurred for routing to path $p_{n0}$ from DU-$n$ to CU. Each split configuration has to satisfy the respective delay requirement (Table \ref{table:splits}):
\vspace{-1mm}
\begin{equation} \label{eq:delay}
x_{on} d_{p_{n0}} \leq d_o^{\text{max}}, \ \ \forall o, \forall n \in \mathcal{N}.
\end{equation}
\subsection{Objective Function}
We aim to minimize the total network cost consisting of the computational cost and routing cost. The computational cost of DU-$n$ is:
\vspace{-1mm}
\begin{align} \label{eq:cost-du}
V_n(\bm{x}_n) = \alpha_n + \beta_n \lambda_{n} \sum_{o=0}^{3}\rho_{o}^\text{d} x_{on}.
\end{align} 
We also have a computing cost for CU:
\vspace{-1mm}
\begin{align} \label{eq:cost-cu}
V_{0}(\bm{x}) = \sum_{n \in \mathcal{N}} \sum_{o=0}^3 x_{on} (\alpha_0 + \lambda_{n} \beta_0 \rho_{o}^\text{c} ).
\end{align} 
Then, the cost to route data flows from DU-$n$ to CU is:
\vspace{-1mm}
\begin{align} \label{eq:cost-route}
U_{n0}(\bm{x}_n) =  \zeta_{p_{n0}} r_n (\bm{x})
\end{align} 
Finally, we have the total vRAN cost as:
\vspace{-1mm}
\begin{align} \label{eq: total-cost}
J(\bm{x}) &= \sum_{n \in \mathcal{N}} \Big(  V_n(\bm{x}_n) + U_{n0}(\bm{x}_n) \Big) + V_{0}(\bm{x}), \end{align}
which leads to the following problem:
\vspace{-1mm}
\begin{align}
\mathbb{P}: \,\,\,\, & \underset{\bm{x} \in \mathcal{X}}{\text{minimize}} \   J(\bm{x})  \notag \\
& \text{s.t} \ \  
 (\ref{eq:computing1}) - (\ref{eq:delay}) \notag
\end{align}
The formulated problem $\mathbb{P}$ is an NP-hard \footnote{It is from the reduction of the Multidimensional Multiple-choice Knapsack Problem (MMKP), see also \cite{vranmec_andres}.}.

%
%

%% file: Algorithms.tex
\section{Solution Approach: DRLT-vRAN} \label{sec:solution}
\vspace{-1mm}
Inspired by \cite{neural_bello, vnf_drl_solozabal}, we aim to solve vRAN split problem with deep RL (for training) and sampling technique with temperature hyperparameter (for inference process), namely \texttt{DRLT-vRAN}. In practice, our approach does not have to know the defined problem in Section \ref{sec:problem}. Our agent interacts with the environment (vRAN) expecting to receive a reward (network cost) and penalization (constraints violation); then learn from this interaction to find the optimal solution.

We utilize Policy Gradient with Lagrangian relaxation method and use neural network architecture that will approximate the policy. Our agent receives input of set of BS functions $\mathcal{F} = \{ \mathcal{F}_n \}_{n=1}^{N} \!\!$ where $\mathcal{F}_n \!=\! \{f_0,f_1,f_2,f_3 \}$ is a set of functions for BS-$n$. In the output, we have $\mathcal{O} \!\!=\!\! \{o_n \}_{n=1}^{N}$ as a set of selected configuration for all BSs. It addresses the split configuration of BS-$n$ with $o_n \! \in \! \{0,1,2,3\}$. 
We use the neural network with weight parameter $\theta$ that infers a policy strategy $\pi_\theta(\mathcal{O}| \mathcal{F}, \theta)$ to deploy the split configuration.
  
\vspace{-2mm}
\subsection{Neural Network Architecture}
\vspace{-1mm}


We use a sequence-to-sequence model with an attention mechanism,
formed by encoder-decoder design, and based on stacked LSTM \cite{seq2seq,attention_bahdanau}.  Since our system has computational and link capacity, the BS input sequence order affects the solution. Hence, we also draw a batch of $B$ i.i.d samples with different sequence order when training our model. Additionally, the attention gives information on how strongly the element of a sequence is correlated to each other; hence it allows to capture the characteristic of BSs.  

Our neural network infers a solution policy strategy to deploy the function split configuration for all BS, given a sequence of BSs as an input $\mathcal{F} = \{\mathcal{F}_1, ...., \mathcal{F}_N \}$. The encoder read the entire input sequence to a fixed-length vector.  Then, the decoder decides the functional split configuration of a BS at each step from an output function based on its own previous state combined with an attention over the encoder hidden states \cite{attention_bahdanau}. The decoder network hidden state is defined with a function: $\bm{h}_t = f(\bm{h}_{t-1}, \bm{\bar{h}}_{t-1}, \bm{c}_t)$. 
Our model also needs to derive a context vector $\bm{c}_t$ that captures relevant source information at each step $t$ that helps predicting the current deployed split configuration. The main idea is to use attention where the context vector $\bm{c}_t$ takes consideration of all the hidden states of the encoder and the alignment vector $\bm{a}_{t}$ as: $\bm{c}_t = \sum_{k \in \mathcal{F}} \bm{a}_{tk} \bm{\bar{h}}_k.$ The alignment vector $\bm{a}_{t}$ has an equal size to the number of step on the source side. It can be calculated by comparing the current target hidden state $\bm{h}_t$ with each source hidden state $\bm{\bar{h}}_k$, hence: $ \bm{a}_{tk} = \texttt{softmax}(\texttt{score}(\bm{h}_t,\bm{\bar{h}}_k)) $, where the score function is defined from Bahdanau's additive style as: $
\texttt{score}(\bm{h}_t,\bm{\bar{h}}_k) = \bm{v}_a^{\top} (\tanh(\bm{w}_1 \bm{h}_t +  \bm{w}_2 \bm{\bar{h}}_k )), $ with $\bm{v}_a^{\top} \in \mathbb{R}^{n}, \bm{w}_1 \!\in\! \mathbb{R}^{n \times n} $ and $\bm{w}_2 \!\in\! \mathbb{R}^{n \times 2n}$ are the weight matrices, and $n$ is the number of neural network layers.

\vspace{-2mm}
\subsection{Policy Gradient with Constraints}
\vspace{-1mm}
\begin{algorithm}[t!]  \caption{\texttt{DRLT-vRAN} Training}
	\SetAlgoLined
	\DontPrintSemicolon
	\KwInitialize{ $T$(Num of epoch), Agent and critic (baseline) networks with  params $\theta$ and $\theta_v. \;$} 
	\Repeat{ T} 
	{
		$\mathcal{F}^i \sim $ \text{SampleInput} $(\mathcal{F})$ $\forall i \in \{1,...,B \}$. \;
		$\mathcal{O}^i \sim $ SampleSolution $(\pi_\theta(.|\mathcal{F}))$ $\forall i \in \{1,...,B \}$. \;
		$b^j \leftarrow b_{\theta_v} (\mathcal{F}^j)$ $\forall j \in \{1,...,B\}$ \;
		Compute $L(\mathcal{O}^i | \mathcal{F}^i)$ \;
		$g_\theta \leftarrow \nabla_\theta J_L^\pi(\theta)$ from Eq. \eqref{eq:lag_grads} \;
		Compute $\mathcal{L}(\theta_v)$ from Eq. \eqref{eq:aux_mse} \;
		$\theta \leftarrow$ Adam($\theta, g_\theta$) $\%$ Run Adam algorithm \;
		$\theta_v \leftarrow$ Adam($\theta_v, \mathcal{L}(\theta_v)$) $\%$ Run Adam algorithm \;
	}
	\Return $\theta, \theta_v$
	\;
\end{algorithm}
A Policy gradient method is applied to learn the parameters of the stochastic policy $\pi_\theta(\mathcal{O}| \mathcal{F}, \theta)$. It predicts the split configuration that minimizes the total cost by assigning a high probability to $o_n $ for having a lower cost and a low probability for a higher cost. Our neural network uses the chain rule for factorizing the output probability:
\vspace{-1mm}
\begin{align}
	\pi_\theta(\mathcal{O}| \mathcal{F}, \theta) = \prod_{n=1}^{N} \pi_\theta(o_n| o_{(<n)}, \mathcal{F}_n).
\end{align}
We define the objective of $\mathbb{P}$ as an expected reward that is obtained for every vector of weight $\theta$. Hence, the expected cost $J$ in associated with the selected configuration given by BS-$n$ functions is:
\vspace{-1mm}
\begin{align}
	J^\pi(\theta|\mathcal{F}_n) = \underset{o_n \sim \pi(.|\mathcal{F}_n) }{\mathbb{E}} [ J(o_n) ],
\end{align}
and we have the expected of total operating cost of all BS:
\vspace{-1mm}
\begin{align} \label{eq:total_cost_theta}
J^\pi(\theta) = \underset{o_n \sim \mathcal{O} }{\mathbb{E}} [ J(\theta|\mathcal{O}) ].
\end{align}
\vspace{-0.5mm}
Our system model has constraints of delay requirements, computational, and link capacities; hence, we define $J_C^\pi(\theta)$ as a function of constraint dissatisfaction to capture the penalization that the environment returns for violating the constraint requirements. 
%
%
Our original problem turns to a primal problem:
\vspace{-3mm}
\begin{align} 
\mathbb{P}_{1\text{P}}: \ \underset{\pi \sim \Pi }{\text{min}} \  J^\pi(\theta); \ \ \text{s.t.} \ J_C^\pi(\theta) \leq 0. \notag
\end{align}
Then, we reformulate $\mathbb{P}_{1\text{P}}$ to unconstraied problem with Lagrange relaxation and penalize the unfeasible solution because of constrains violation, following approach in \cite{reward_constraint,vnf_drl_solozabal}. Hence, we have the dual function:
\vspace{-1.5mm}
\begin{align} 
	\label{eq:dual_function}
	g(\bm{\mu}) = \underset{\theta }{\text{min}} \  J_L^\pi(\bm{\mu},\theta) &= \underset{\theta }{\text{min}} \  J^\pi(\theta) + \bm{\mu} J_C^\pi(\theta) \notag \\
	&=\underset{\theta }{\text{min}} \  J^\pi(\theta) + J_\zeta^\pi(\xi)
\end{align}
where $J_L^\pi(\bm{\mu},\theta)$ and $J_\zeta^\pi(\xi)$ are the Lagrange objective function and the expected penalization, respectively. The expected penalization is defined with a weighed sum of all expectations of constraint dissatisfactions. Finally, we define the dual problem:
\begin{align}\label{eq:dual_problem}
\mathbb{P}_{1\text{D}}: \ \underset{\bm{\mu} }{\text{max}} \  g(\bm{\mu}). \notag
\end{align}
The weight parameter $\theta$ that optimizes the objective can be computed by using Monte-Carlo Policy Gradient (\texttt{REINFORCE}) with baseline estimation:
\begin{align}
	\theta_{k+1} = \theta_{k} + \alpha 	 \nabla_\theta J_L^\pi(\theta),
\end{align} 
where the gradient is calculated with log-likelihood method:
\begin{align}
\nabla_\theta J_L^\pi(\theta) = \underset{\mathcal{O} \sim \pi_\theta(.|\mathcal{F}) }{\mathbb{E}} [ L(\mathcal{O}|\mathcal{F}) \ \nabla_\theta \log \pi_\theta(\mathcal{O}|\mathcal{F}) ].
\end{align} 
$L(\mathcal{O}|\mathcal{F})$ is the total cost with penalization. This cost consists of the total cost and the weighted sum of penalization: 
	 $L(\mathcal{O}|\mathcal{F}) = J(\mathcal{O}|\mathcal{F}) + \xi (\mathcal{O}|\mathcal{F}) $, where
$J(\mathcal{O}|\mathcal{F})$ is the total operating cost in each iteration and $\xi (\mathcal{O}|\mathcal{F}) = \bm{\mu} C(\mathcal{O}|\mathcal{F})$ is the weighted sum of constraint dissatisfaction of $C(\mathcal{O}|\mathcal{F})$. Next, we use Monte-Carlo sampling to approximate the gradient by drawing  $B$ i.i.d samples $\mathcal{F}^1,...,\mathcal{F}^B \sim \mathcal{F}$; hence, the gradient turns to:
\begin{align} \label{eq:lag_grads}
\!	\nabla_{\!\theta} J_L^\pi(\theta) \! \approx \! \frac{1}{B} \! \sum_{i=1}^{B} \! \! \Big(\! L(\mathcal{O}^i | \mathcal{F}^i) \! - \! b_{\theta_v}(\mathcal{F}^i)\! \Big) \! \nabla_{\!\theta} \! \log \! \pi_\theta(\mathcal{O}^i | \mathcal{F}^i). \!\!
\end{align} 
The baseline choice can be from an exponential moving average of the reward over the time that captures the improving policy in training. Although it succeeds in the Christofides algorithm, it does not perform well because it can not differentiate between different input \cite{neural_bello}. To this end, we use a parametric baseline $b_{\theta_v}$ to estimate the expected total cost with penalization $\mathbb{E}_{\mathcal{O} \sim \pi(.|\mathcal{F})} L(\mathcal{O}|\mathcal{F})$ that typically improves learning performance. We train the baseline in an auxiliary network built from an LSTM encoder connected to a multilayer perceptron output layer similar to  \cite{vnf_drl_solozabal}. The auxiliary network (parameterized by $\theta_v$) that learns the expected cost with penalization by the current policy $\pi_\theta$ from given input $\mathcal{F}$, is trained with stochastic gradient descent. It employs a mean squared error (MSE) objective, calculated from its prediction $b_{\theta_v}$ and the actual cost with penalization, and sampled by the most recent policy (obtained from the environment). Finally, we formulate the auxiliary objective:
\vspace{-1mm}
\begin{align} \label{eq:aux_mse}
	\mathcal{L}(\theta_v) = \frac{1}{B} \sum_{i=1}^{B} \left\| b_{\theta_v}(\mathcal{F}^i) - L(\mathcal{O}^i | \mathcal{F}^i) \right\|_2^2,
\end{align}
and summarize our approach in Algorithm 1. 

\vspace{-3mm}
\subsection{Searching Strategy}
\vspace{-1mm}
We use the \texttt{DRLT-vRAN} training model to predict the solution. We utilize a search strategy in the test, particularly a sampling technique with a temperature hyperparameter \cite{neural_bello}. It considers multiple candidate solutions, then infers the best solution. The approach is to sample candidate solutions from stochastic policy, then select the split configuration with the lowest total cost. A temperature hyperparameter controls the diversity of the sampling to attain an improvement in finding the best solution. The detailed algorithm is described in \cite{neural_bello}.

%% file: Simulations.tex

\begin{figure}[t]
	\centering
	\begin{subfigure}[t]{.23\textwidth}
		\centering
		\includegraphics[width=\textwidth]{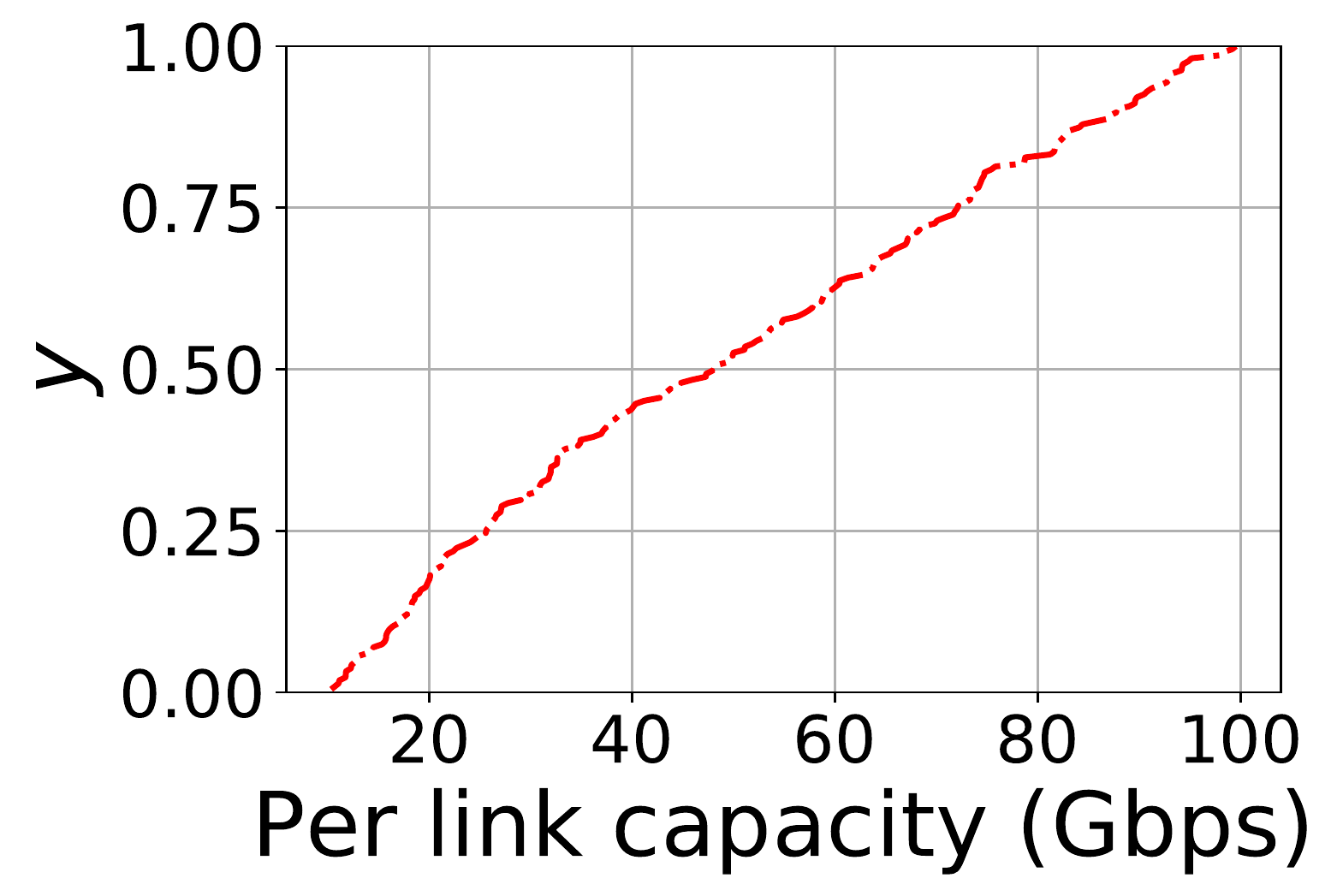}
			\vspace{-5mm}	
		\small\caption{\small }
	\end{subfigure}
	\begin{subfigure}[t]{.23\textwidth}
		\centering
		\includegraphics[width=\textwidth]{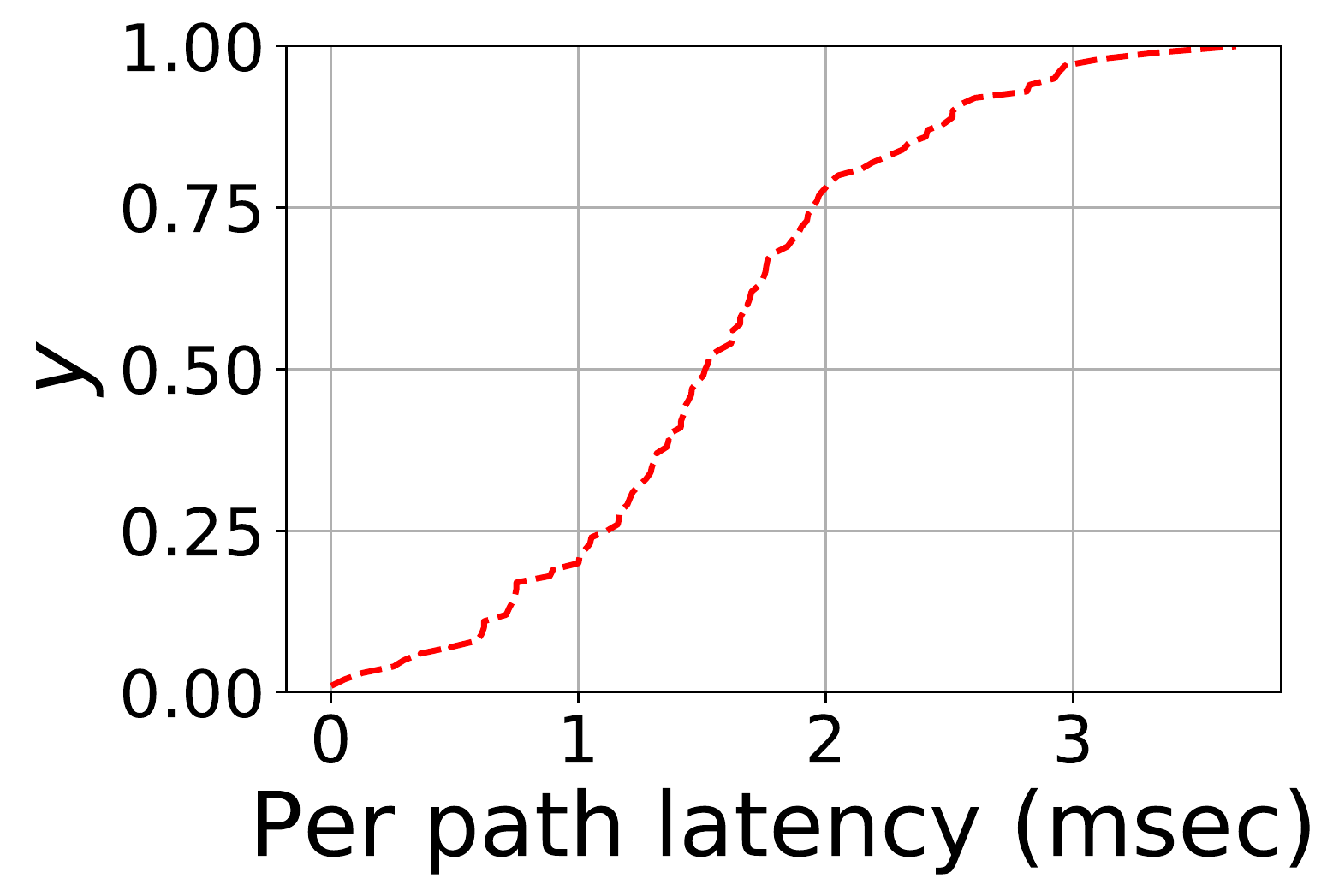}
			\vspace{-5mm}	
		\small\caption{\small }
	\end{subfigure}
	\caption{\small \textbf{RAN dist.} eCDF of (a) per-link capacity; (b) per-path latency.}
	\label{fig:ran}	
	\label{fig:ran_params}
		\vspace{-5mm}		
\end{figure}


\begin{figure}[t]
	\centering
	\includegraphics[width=.47\textwidth]{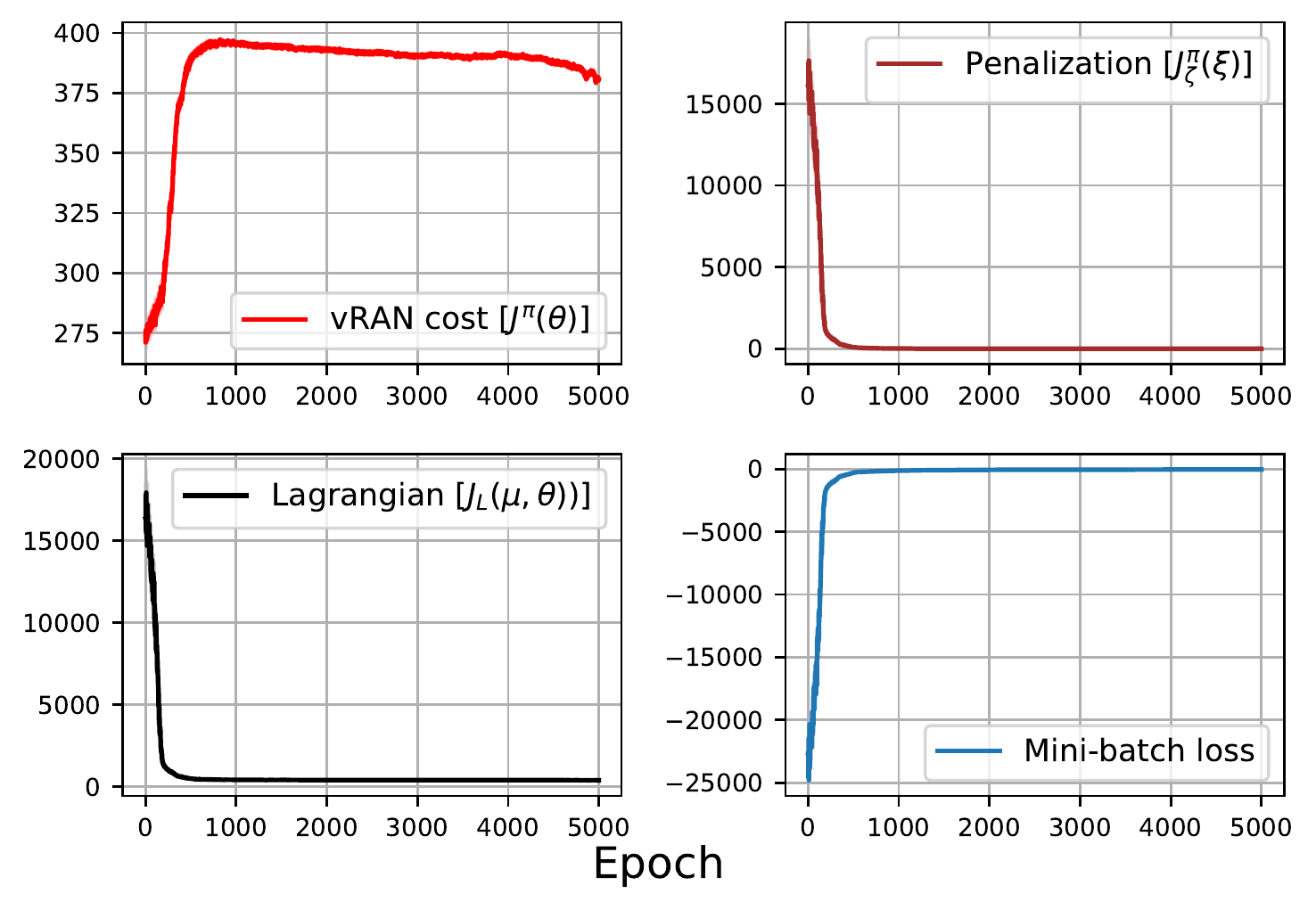}
	\vspace{-3mm}			
	\caption{\small \textbf{\texttt{DRLT-vRAN} training.} Results of training R1. }
	\label{fig:res_train}
	\vspace{-5mm}	
\end{figure}

\section{Results and Discussion} \label{sec:results}
\vspace{-1mm}
In this section, we conduct several experiments to evaluate \texttt{DRLT-vRAN} using a synthetic network. We aim to examine: \textit{(i)} the behaviour of \texttt{DRLT-vRAN} during training process, \textit{(ii)} the accuracy of \texttt{DRLT-vRAN} and the searching strategy to infer the best solution over feasible solutions, w.r.t \textit{(iii)} different traffic loads and routing costs.

\vspace{-3mm}
\subsection{Environment \& Experiment Setup}
\vspace{-1mm}
We use a synthetic RAN (R1) to evaluate \texttt{DRLT-vRAN}. R1 is generated using Waxman algorithm \cite{waxman} with parameters such as link probability ($\alpha$) and edge length control ($\beta$). These respective parameters $(\alpha,\beta)$ are set to $(0.5, 0.1)$. R1 has 100 nodes consisting of 1 CU and 99 DUs. R1 differs in parameters, e.g., location, link capacity, weighted link, delay. The link capacity and path delay vary to $100$ Gbps and $3658.61 \ \mu s$, respectively. Finally, we visualize our RAN distributions with eCDF in Fig. \ref{fig:ran_params}.

In this experiment, all system parameters correspond to testbed measurements of previous studies \cite{vranmec_andres,fluidran_andres,vran_murti1,vran_murti2,cost_vm}. We assume a high load scenario $\lambda_{n} = 150$ Mbps for every DU\footnote{This setting is based on 1 user/TTI, $2 \times 2$ MIMO, 20 Mhz (100 PRB), 2 TBs of 75376 bits/subframe and IP MTU 1500B.}. We use an Intel Haswell i7-4770 3.40GHz CPU as the \textit{reference core}, and set the maximum computing capacity to 75 RCs for CU and 7.5 RCs for each DU. Each split configuration $o \in \{ 0,1,2,3 \}$ inccurs computational load $\rho_{o}^{\text{d}} = \{ 0.05, 0.04, 0.00325, 0\}$ RCs per Mbps for DU and $\rho_{o}^{\text{c}} = \{0, 0.001, 0.00175, 0.05 \} $ RCs per Mbps for CU, respectively. The VM instantiation cost of CU is a half of DU $(\alpha_0 = \alpha_n/2)$ and the processing cost is set to $\beta_0 = 0.017 \beta_n$. Finally, the routing cost per path depends the cost of per selected link where it comes from randomly generated $ \left[ 0,1 \right] $ per Gbps for each link\footnote{A link with a routing cost of 1 monetary unit per Mbps means having the same cost as a DU computing cost. We consider a cheap routing cost with a range of $0.0001 - 0.001\times$ of DU computing cost for each link.}. 

Our learning rate is initially set to $0.0001$ (agent) and $0.005$ (baseline) with the batch size: 128. Our neural network has the number of layers, hidden dimension, and embedding size with $1, 32$, and $ 32$, respectively. We scale all weighted paths and traffic loads randomly with uniform distribution $[0,1]$ as in \cite{neural_bello}. Then, we generate three models (\textit{RL-pretained}) as an output of our training with 15000 epochs each. The training is performed with Tensorflow 1.15.3 and Python 3.7.4.

\vspace{-2mm}
\subsection{Training Analysis}
\vspace{-1mm}

Fig. \ref{fig:res_train} visualizes the training behaviour of \texttt{DRLT-vRAN}. We found some values for the cost of penalization at the beginning of the training. This behavior occurs because \texttt{DRLT-vRAN} tries to find the solution but it violates the constraint sets (e.g., latency, bandwidth, computation); hence it receives additional cost for penalization. It also shows that there is a significant difference in the vRAN cost ($J^\pi(\theta)$) compared to the Lagrangian cost ($J_L(\mu,\theta)$). This fact describes that our agent receives a high penalization from the environment R1. It is proven that R1 has several large path delays and small link capacities. Our agent learns to improve its policy by focusing on penalization first and then correcting its weights via stochastic gradient descent. It explains how the vRAN cost increases (focusing on reducing penalization cost) and adjusting as the training process goes. Then, we have the Lagrangian cost ($J_L(\mu,\theta)$) that considers both vRAN and penalization cost. It describes how our agent tries to minimize the primal problem $\mathbb{P}_{\text{1P}}$ through the dual problem $\mathbb{P}_{\text{1D}}$. When our agent finally dismisses the penalization cost, the Lagrangian cost becomes equal to the vRAN cost. The minibatch loss decreases to near zero after several epochs as our agent improves the policy. Finally, our agent learns until it finds the local minima, or saddle point, or ends by reaching the number of epochs.

\vspace{-1mm}
\textbf{Findings:} \textit{1) Our agent improves its policy by focusing on the penalization; then, it adjusts its weight as the training goes. 2) The Lagrangian cost is decreasing, then turns equal to the vRAN cost as all constraints are satisfied.}
\vspace{-1mm}

\begin{figure}[t!]
	\centering
	\includegraphics[width=.46\textwidth]{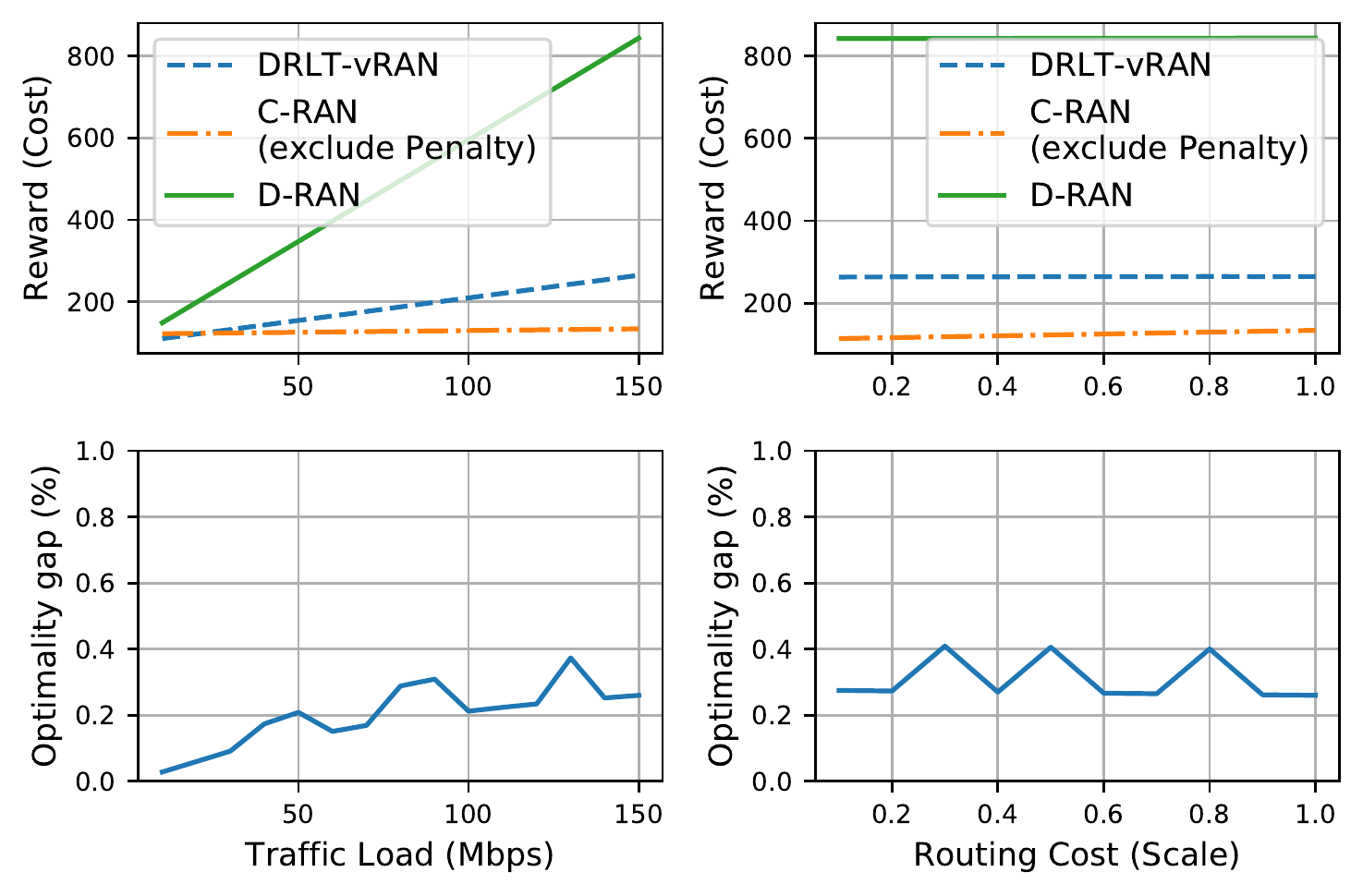}
	\vspace{-3mm}		
	\caption{\small \textbf{\texttt{DRLT-vRAN} test.} Results of altering traffic load and routing cost to the total of vRAN cost and optimality gap over R1, and the incurred total system cost for implementing \texttt{DRLT-vRAN} compared to D-RAN and C-RAN.}
	\label{fig:test_n2}
	\label{fig:res_test}	
	\vspace{-5mm}
\end{figure}

\vspace{-2mm}
\subsection{Impact of Load and Routing Cost}
\vspace{-1mm}

This part studies the impact of altering system parameters, e.g., cost and traffic load, to \texttt{DRLT-vRAN}, then compared it to the optimal solution\footnote{We use the term \emph{optimality gap} to define our solution's error compared to the optimal value obtained from Python-MIP Solver.}. This evaluation is conducted using pre-trained models as results of \texttt{DRLT-vRAN} training and search strategy for the inference process. Firstly, we consider traffic loads within a range $\lambda_{n} = [10,150]$ Mbps and use the original routing cost of R1. Then, we change the scale of routing cost, i.e., increasing or decreasing the leasing agreement's price, maintenance, etc., within a range $[0.1,1]$ and using default traffic load $\lambda_{n}=150$ Mbps. We also benchmark to two extremes of RAN setups, fully D-RAN and C-RAN\footnote{We practically can not implement C-RAN because our RAN does not meet the constraint requirements of delay, bandwidth, and CU capacity, to deploy C-RAN. The presented C-RAN in Figs. \ref{fig:res_test} is just for benchmarking, hence we also do not consider the penalization cost (constrains violation) for this case.}.

Fig. \ref{fig:res_test} depicts that \texttt{DRLT-vRAN} is still capable to find the solution that is very close to the optimal solution ($<0.4 \%$) for both scenarios: increase of the traffic load and routing cost. Although the optimality gap firstly rises in line with the traffic load, it diminishes as the traffic load is getting high ($>100$ Mbps), with a less than $0.4\%$ trend. Further, there is no significant impact on the optimality gap for increasing the routing cost (stable within a range $0.2 -  0.4 \%$). It is worth noting that employing a sampling technique with a temperature hyperparameter during the inference process allows selecting the index with the largest probability resulting in improvement in finding the best solution \cite{neural_bello}. Hence, there is also a tightened optimality gap at some points. Fig \ref{fig:res_test} also shows that  \texttt{DRLT-vRAN} is provably cost-effective compared to D-RAN with $320\%$of cost-saving, yet it still has a higher cost to C-RAN. However, it worth noting that  \texttt{DRLT-vRAN} considers penalization (constrains violation), hence  \texttt{DRLT-vRAN} can not push its functions more centralized. Since C-RAN violates many constraints, it is useful only for a reference. It shows that \texttt{DRLT-vRAN} cost is not far from C-RAN. Instead, at the low traffic load, \texttt{DRLT-vRAN} is more efficient than C-RAN. Lastly, the increase of C-RAN cost to the rise of routing cost is higher compared to  \texttt{DRLT-vRAN}. 

\vspace{-0.5mm}
\textbf{Findings:} \textit{1) \texttt{DRLT-vRAN} solves the problem with a very small optimality gap of smaller than $< 0.4 \%$ even by altering the traffic load and routing cost. 2) \texttt{DRLT-vRAN} is cost-effective and can achieve 320$\%$ of cost saving compared to D-RAN. 3) The increase in traffic load gives the most impact to D-RAN while the routing cost affects C-RAN at the most. }
\vspace{-0.5mm}

%% file: Conclusions.tex
\section{Conclusion} \label{sec:conclusion}
\vspace{-1mm}
In this paper, we have proposed a policy gradient based deep reinforcement learning method to solve the vRAN function split problem with minimal handcrafted engineering. We have proposed a learning algorithm that learns a stochastic policy to decide where the BS functions are hosted, either at DU or CU. We also have considered vRAN environment constraints (server \& link capacity and delay requirements) and penalization for violation.  A search strategy has been utilized to infer the best solution in the test. We have evaluated our proposed solution in a synthetic RAN simulation set up. The results have shown that our approach successfully learns function splitting decision with less than $0.4\%$ optimality gap and outperforms D-RAN.